\documentclass[aps,prl,
twocolumn,
superscriptaddress,
amsfonts,amssymb,amsmath,
nofootinbib,
twoside,
letterpaper,
]{revtex4}

\usepackage{bm,color}

\begin{document}

\title{Universal features of the defect-induced damping of lattice vibrations}

\author{A. Cano}
\email{andres@iaa.es}
\affiliation{Instituto de Astrof\' \i sica de Andaluc\' \i a CSIC, PO Box 3004, E-18080 Granada, Spain}
\author{A. P. Levanyuk}
\email{levanyuk@uam.es}
\affiliation{\mbox{Departamento de F\'\i sica de la Materia Condensada C-III, Universidad Aut\'onoma de Madrid, E-28049 Madrid, Spain
}
}
\author{S. A. Minyukov}
\email{minyukov@mail.ru}
\affiliation{Institute of Crystallography, Russian Academy of Sciences, Leninskii Prospect 59, Moscow 117333, Russia }

\date{\today}

\begin{abstract}
It is shown that any defect gives an Ohmic contribution to the damping of any normal mode of the crystal lattice with nonzero wavevector which does not vanish at zero temperature. This explains the large phason damping observed at low temperatures in incommensurate phases, and might be a key factor to understand the linear-in-$T$ specific heat observed in a number of real dielectrics at low enough temperatures.
\end{abstract}

\maketitle

It is well known since long ago that the defect-induced damping of the normal modes of the crystal lattice is a key factor to understand the real dielectric losses \cite{Gurevich91}, attenuation of sound \cite{Truell_Elbaum_Chick}, spin lattice relaxation rates, etc. This is especially clear at low temperatures, where the real damping cannot have its origin in the well studied effects of anharmonism: a strong dependence on temperature $\sim T^{5}\div T^{7}$ is predicted for the phonon attenuation \cite{Reissland,Levanyuk97a} which is incompatible with the low-temperature data \cite{Truell_Elbaum_Chick}. Recently it has been argued that, at low temperatures, the defect-induced damping can play an even more far reaching role as long as it may be the reason of the deviations from the Debye law observed in the low-temperature specific heat of real dielectrics \cite{Cano_a,Cano_b}. In this paper we show that \emph{the defect-induced damping is unexpectedly universal} in the sense that any defect gives an Ohmic contribution to the damping of any normal mode of the crystal lattice with nonzero wavevector. This feature reinforces the candidature of the defect-induced damping as the main reason for the aforementioned deviations from the Debye law.

In accordance with previous results restricted to a $\mathbf k=0$ polar mode, the presence of a random distribution of charge defects makes the damping of this mode to be Ohmic at low frequencies, i.e., without frequency dispersion \cite{Vinogradov61}. This can be easily understood as follows. When applying a homogeneous electric field, i.e., the force conjugated to the homogeneous polarization, the charge defects do displace. These displacements are accompanied with (local) displacements of the corresponding lattice. So, when the field varies in time, the motion of the charges provokes the radiation of acoustic waves whose intensity is $\propto \omega^2$ (not $\propto \omega^4$ as it could be inferred from Ref. \cite{Landau_FM} for the case of local deformations). This radiation gives rise to dielectric losses. In Ref. \cite{Cano_a} it was shown that, in the case of a correlated density of charge defects, this result is also valid for most of the modes in a polar branch. In the following we shall show that, for any lattice mode with nonzero wavevector, any type of defect will also induce the same type of damping.

For the sake of illustration, let us consider a polar branch as in the previous papers. It is worth mentioning, first of all, that the presence of defects means that there is a redistribution of charges with respect to that in the ideal lattice. This allows us to model any defect as a set of electric multipoles. As we shall see, an inhomogeneous electric field varying in time has the above effect on any electric multipole. That is, the field induces the radiation of acoustic waves through the multipoles and, consequently, gives rise to losses. To be specific, let us consider randomly distributed dipolar defects. The charge density associated with these defects can be written as
\begin{align}
\rho(\mathbf r ) 
&=\sum_{i} ({\mathbf p_i\cdot \nabla })\delta (\mathbf r - \mathbf r_i ),
\label{random_dipoles}\end{align}
where $\mathbf p_i = q \mathbf d_i$ and $\mathbf r_i$ are (random) vectors representing the dipolar momenta of the defects and their positions, respectively. (For the sake of simplicity, we further consider that the dipoles are oriented randomly.)

The frequency dependence of the damping constant $\gamma$ of a polar mode can be obtained, in the low-frequency regime we are interested in, from the frequency dependence of the imaginary part of the corresponding susceptibility $\chi (\omega,{\mathbf k})$: $\gamma(\omega\to 0 , {\mathbf k})\propto\omega^{-1} \chi''(\omega\to 0 , {\mathbf k})$. To compute such a susceptibility, we first take into account that, in Fourier space, the polarization can be written as
\begin{align}
\mathbf P (\omega, {\mathbf k}) =& 
\chi_0(\omega,{\mathbf k})\mathbf E(\omega,{\mathbf k}) 
\nonumber \\&
+ {1\over V }\iint \langle \rho ({\mathbf r})\mathbf u ({\mathbf r},t)\rangle
e^{-i ( {\mathbf k}\cdot {\mathbf r}- \omega t) } d{\mathbf r} dt.
\label{PolaP}\end{align}
Here $\chi_0$ is the (scalar) susceptibility in absence of defects (the medium is assumed to be isotropic), ${\mathbf E}$ is the (inhomogeneous) electric field, $V$ is the volume of the system, $\rho$ is the charge distribution due to the defects, $\mathbf u$ is the acoustic displacement vector, and $\langle \dots \rangle$ denotes statistical average. The equation of motion for the displacement vector is 
\begin{align}
\ddot {\mathbf u} = c_t^2 \nabla^2 {\mathbf u}+ (c_l^2- c_t^2)\nabla (\nabla \cdot {\mathbf u}) + \rho{\mathbf E}/\mu,
\end{align}
where $c_l$ and $c_t$ are the longitudinal and the transversal velocities of sound, respectively, and $\mu $ the density of the medium. [The last term in this equation is just the force that the electric field exerts on the lattice due to the presence of the (charges) defects.] In Fourier space, the $i$-th component of this displacement vector then can be written as \begin{align}
u_i(\omega,{\mathbf k})=
- \Lambda_{ij}^{-1}(\omega,{\mathbf k}) {V\over \mu}
\int {d \mathbf k' \over {(2\pi)^3}} 
\rho({\mathbf k}-{\mathbf k'})E_j(\omega,{\mathbf k'}), 
\label{u}\end{align}
where \begin{align}
\Lambda_{ij}^{-1}(\omega,{\mathbf k})
={(\omega^2 - c_l^2 k^2)\delta_{ij}+(c_l^2 - c_t^2 )k_ik_j
\over (\omega^2 - c_l^2 k^2)(\omega^2 - c_t^2 k^2)}.
\end{align}
Substituting Eq. \eqref{u} into Eq. \eqref{PolaP} we find that 
\begin{align}
P_i (\omega, {\mathbf k}) = &
\chi_0(\omega,{\mathbf k})E_i(\omega,{\mathbf k}) 
\nonumber \\
&- {V\over \mu}\int {d \mathbf k' \over {(2\pi)^3}} 
K({\mathbf k}-{\mathbf k'})
\Lambda_{ij}^{-1}(\omega,{\mathbf k'})
E_j(\omega,{\mathbf k}),
\label{P_E}\end{align}
where $K({\mathbf k})$ is defined from the ``spectral density'' of the charge distribution:
$\langle \rho({\mathbf k})\rho({\mathbf k'}) \rangle = K({\mathbf k})V^{-1}
\delta({\mathbf k}+{\mathbf k'})$ \cite{note_delta}. 
In accordance with Eq. \eqref{random_dipoles}, this magnitude is \begin{align}
K({\mathbf k}) = N p^2 k^2 /(3V)
\label{}\end{align} 
in our case, where $N$ is the density of defects.

As long as we are considering an isotropic medium, after integrating over wavevectors we find that the contribution of the non-diagonal terms of $\Lambda_{ij}^{-1}$ in Eq. \eqref{P_E} is zero. This means that the contribution to the susceptibility due to the defects can be written as
\begin{align}
\chi_\text{def} (\omega,{\mathbf k})= -{V\over \mu}\int {d \mathbf k' \over {(2\pi)^3}} 
K({\mathbf k}-{\mathbf k'})\Lambda_{ii}^{-1}(\omega,{\mathbf k'}) 
\label{chiii}\end{align}
(here double subscript does not imply summation). 
The main contribution to the integral in Eq. \eqref{chiii} comes from the poles of $\Lambda_{ii}^{-1}(\omega,{\mathbf k'})$. We are interested in the case of small frequencies. Taking into account that the poles of $\Lambda_{ii}^{-1}(\omega,{\mathbf k'})$ are such that $k'\sim \omega/c$ ($c_l,c_t\sim c$), in the integrand of Eq. \eqref{chiii} we can take
\begin{align}
K({\mathbf k} - {\mathbf k}')\approx N {p^2 k^2 }/(3V),
\end{align}
for large wavevectors ($k\gg \omega/c$). We then have 
\begin{align}
\chi_\text{def}'' (\omega \to 0,k\gg \omega/c) \sim {N\over N_\text{at}} {p^2 k^2\over e^2} {\omega\over \omega_D},
\end{align}
where $N_\text{at}=l_\text{at}^{-3}$ is the atomic density and $\omega _D = c /l_\text{at}$ is the Debye frequency. In this expression we see that, though the damping vanishes for wavevectors tending to zero, it is quite significant for wavevectors close to the boundary of the corresponding Brillouin zone (where the factor $p k$ is $\sim q$). 

At this point, it is worth mentioning that, at large enough wavevectors, the modes of any branch are effectively polar (strictly speaking, the label polar or nonpolar refers to $\mathbf k=0$). Therefore, the above considered mechanism for damping gives a contribution for either optic or acoustic modes. In the latter case, the polarization may have its origin in either ions or electrons. It is worth noticing that, in any case, the strength of the damping is the same for $\omega \to 0$ [the only parameters entering in Eq. \eqref{chiii} are elastic constants and the multipole moment].

It is also worth mentioning that similar results are obtained not only for higher order electric multipoles, what can be easily seen by following similar arguments, but also in the case of 
multipoles of different nature. In fact, the change in the charge distribution of a crystal is only one of the several aspects of the influence of defects. Among these other aspects, much attention has been paid to the elastic one for which, the modelling of the defects as ``force multipoles,'' has been proved to be quite useful (see, e.g., Ref. \cite{Teodosiu}). Inhomogeneous strains and stresses then play a role analogous to that of the inhomogeneous polarization and electric fields considered above. 

Defects can also be considered as more general multipoles, say ``optic multipoles.'' This became apparent when studying phase transitions where, acting as a (local) field conjugated with the corresponding order parameter, defects may play a key role (see, e.g., Ref. \cite{Levanyuk_Sigov}). In view of this, the above mechanisms may explain, for instance, the unexpected phason damping observed at low temperatures (see, e.g., Ref. \cite{PhasonDamping}). The theory of anharmonic crystals cannot explain this damping as long as this theory predicts a very strong dependence on temperature \cite{Levanyuk97a}. But taking into account that, in terms of the normal phase, phasons are optic or acoustic vibrations with large wavevectors (close to that of the corresponding incommensurate structure), it is reasonable to expect a significant contribution due to the above mechanism of damping. 

As an interesting consequence of the above results, it is the fact that a damping with such features provides a natural explanation for the, otherwise difficult to explain, (quasi)linear-in-$T$ contributions observed in the low-temperature specific heat of a number of incommensurate dielectrics and charge-density-wave systems \cite{Dahlhauser86}. The mechanism of damping we have discussed naturally explains also the non-Debye specific heat observed in a number of ferroelectrics some time ago \cite{Lowless76}, which was experimetally linked to the presence of defects \cite{Villar86}. Let us detail this point to some extent.

In principle, the low-temperature properties of dielectrics are determined by the corresponding density of (phonon) states at frequencies tending to zero (see, e.g., Ref. \cite{Landau_SP}). In real dielectrics, i.e., dielectrics containing defects, the general features of such a density of states, however, are normally unknown. Whether this density of states remains finite at $\omega =0$ is a question of particular interest: In this case, the crystal will exhibit, among other characteristic features, a linear-in-$T$ specific heat at low enough temperatures; what is normally considered as a fingerprint of glasses (see, e.g., Ref. \cite{Phillips87}). 

When the concentration of defects is small, it seems plausible to study the low-temperature properties of real systems, instead of focusing on the corresponding density of states, by following the approach exposed in Refs. \cite{Cano_a,Cano_b} (see also Ref. \cite{Bulaevskii93}). When doing so, one first considers the phononic scenario of the clean (ideal) crystal and then encapsulates the influence of defects into a defect-induced damping of the corresponding phonons \cite{Cano_a,Cano_b}. The possibility of identifying a reservoir where the energy lost by the damped phonons can be transferred to, without altering the properties of such a reservoir, allows one to further compute the corresponding low-temperature properties within the already developed framework of quantum dissipative systems (see, e.g., Ref. \cite{Weiss}). Long-wavelength acoustic phonons brings about the mentioned reservoir \cite{Cano_a}, and the validity of the approach is guaranteed by virtue of the smallness of the defect concentration ---i.e., the weakness of coupling between the (sub)system of interest and the reservoir. Within this context, to have an Ohmic damping for an harmonic oscillator is equivalent to have a finite density of states at $\omega = 0$ and, though it has not been demonstrated explicitly, it is quite expectable that the same happens when one considers a whole crystal (i.e., not a single oscillator but a set of them). Bearing in mind that the above Ohmic damping is precisely the one we have demonstrated to be ``universal,'' the specific heat of real dielectrics has to be expected to be linear-in-$T$ at low enough temperatures.

A natural question is the relevance of these results for glasses. Strictly speaking, glasses are beyond of the applicability of the approach in Refs. \cite{Cano_a,Cano_b} as long as this is designed to deal with small concentrations of defects only. However, it is quite reasonable to speculate that, behind the low-temperature specific features of glasses, there are significant contributions that may be interpreted as due to the damping of the corresponding vibrational excitations.

In summary, we have shown that \emph{any defect gives an Ohmic contribution to the damping of any lattice vibration with nonzero wavevector}. This contribution does not vanish at zero temperature and, consequently, may be a key point to understand the low-temperature properties of real systems that, despite a small concentration of defects, exhibit universal glass-like features. 

We thank C. Barcel\' o, G. G\' omez-Santos and M.A. Ramos for useful discussions. S.A.M. supported by the Russian Foundation for Basic Researches (RFBR), Grant 05-02-17565.


\begin{references}
\bibitem{note_delta} $\delta (\mathbf k)$ is the Dirac delta function in $\mathbf k$-space: $\int d \mathbf k \delta (\mathbf k) =1$.
\end{references}
\end{document}